\documentstyle[12pt]{article}

\begin{document}
\newcommand{\non}{\nonumber}
\newcommand{\lan}{\langle}
\newcommand{\ran}{\rangle}
\newcommand{\ra}{\rightarrow}
\newcommand{\Lamc}{\Lambda_{c}}
\newcommand{\Lam}{\Lambda}
\newcommand{\nn}{\nonumber}

\begin{flushright}
\vspace{-3.0ex}
    {\sf ADP-02-76/T515} \\
\vspace{-2.0mm} \vspace{5.0ex}
\end{flushright}

\centerline{\large\bf Cabibbo-suppressed non-leptonic decays of
$\Lambda_c$ and final state interaction}

\vspace{1cm}

\centerline{Shao-Long Chen$^{1,2}$, Xin-Heng Guo$^{4}$, Xue-Qian
Li$^{1,2,3}$ and Guo-Li Wang$^{1,2,3}$}

\vspace{0.5cm}
\begin{center}
\small{ 1. CCAST (World Laboratory), P.O.Box 8730, Beijing 100080,
China,

2. Department of Physics, Nankai University, Tianjin, 300071,
China.

3. Institute of Theoretical Physics, Academia Sinica, Beijing
100080, China.

4. Department of Physics and Mathematical Physics, and Special
Research Center for the Subatomic Structure of Matter, Adelaide
University, SA 5005, Australia}

\end{center}

\vspace{0.5cm} \baselineskip 10pt

\begin{center}
\begin{minipage}{11cm}
\noindent Abstract

With the diquark structure of $\Lambda_c$, we investigate the
branching ratios of $\Lambda_c\rightarrow n\pi^+$ and $p\pi^0$.
The results show that without considering the final state
interaction (FSI), the branching ration of $\Lambda_c\rightarrow
p\pi^+$ is only of order $10^{-6}$ whereas taking into account the
FSI effects, this ratio could reach $10^{-4}$ and  is at the same
order as $\Lambda_c\rightarrow n\pi^0$. Concrete values depend on
phenomenological parameters adopted in the calculations. These
branching ratios can be measured in the experiments to come.

\end{minipage}
\end{center}
\vspace{0.5cm}

\baselineskip 22pt \noindent{PACS numbers: 14.20.Lq, 13.30.-a,
12.39.Hg, 12.39.-x}

\vspace{1cm}

\baselineskip 20pt

\noindent{\bf I. Introduction}
\\

In the recent years, there are remarkable progresses in heavy
flavor physics for both theoretical framework and applications. As
long as heavy flavors are involved in the concerned physical
processes, due to an extra symmetry $SU(2)_f\otimes SU(2)_s$,
calculations become simpler and more reliable, and the physics
picture is clearer as well \cite{Isgur}. Based on the
understanding, the heavy quark effective theory (HQET) has been
established \cite{Georgi} and applied to investigate the physical
processes. In the HQET, at the leading order of $1/m_Q$ expansion,
the non-perturbative QCD effects are attributed into one form
factor, i.e. the Isgur-Wise function. With it, one can make
predictions and extract significant information from data, such as
the CKM entries.

Heavy-meson related processes have attracted intensive interests
because a great amount of data about B and D have been accumulated
so far. By contrary, the processes of heavy baryons have not
received as much attention as for mesons. The reason is two-fold.
First, there are not sufficient amount of data on heavy baryons
available yet. Secondly, one has to deal with a three-body problem
theoretically which is much more complicated than the meson case.
On the other side, there are many unsolved problems for baryons,
for example, the lifetime of $\Lambda_b$ is 27\% shorter than that
of B-meson, and it indicates the shortcoming of the spectator
scenario \cite{Bigi}.

The diquark picture was proposed a long time ago and it has been
applied to study many processes involving baryons \cite{Kroll}.
Obviously, for a baryon  containing two heavy and one light
quarks, the two heavy quarks constitute a diquark which serves as
a color-field source for the light  flavor\cite{Wise}. Instead,
for the one-heavy and two-light quark structure, it is suggested
that the two light quarks can also form a light diquark.

In this work we are going to employ the diqaurk picture to study
the Cabibbo-suppressed $\Lambda_c$ decays. The reason for doing
such calculations is that in the near future there will be a great
amount of data on $\Lambda_c$ decays available and probably,
higher statistics for the rare decay modes will also be achieved,
so we can use the data to testify our theory. Moreover, in these
decays the final state interaction (FSI), which is also a very
important subject, is more important than that in the case of
$\Lambda_b$. In this work, we will investigate how much the FSI
effects could change the branching ratios.

We will mainly use the factorization ansatz which is not accurate
for meson decays, as many authors pointed out
\cite{Buras,Shifman,Cheng}. Even though the non-factorizable
effects which are proportional to the color-octet currents may
bring up some errors for the theoretical predictions, later we
will show that they do not affect our qualitative conclusion.
Besides, there are some uncertainties coming from other
approximations and all of them are due to lack of sufficient
knowledge on the non-perturbative QCD.

All the weak effective vertices employed in this work are standard
\cite{Deshpande}. Then we can calculate the weak transition in a
straightforward way up to the hadronic matrix elements. To
evaluate the hadronic matrix elements, one needs to invoke models,
because they are fully governed by the non-perturbative QCD. Since
we are dealing with transitions from $\Lambda_c$ to lighter
baryons $\Lambda,p,n$, etc., we cannot simply use the Isgur-Wise
function which applies only to the transitions between heavy
flavors. In terms of the diqaurk picture, we employ the
wave-functions of baryons which are presented in literature to
carry out calculations of the corresponding hadronic matrix
elements. They finally are attributed into two independent form
factors.

Another important issue is the FSI effects which may bring up
dramatic changes to the predicted branching ratios. As well known
in the meson case, FSI plays a remarkable role and sometimes is
almost crucial \cite{Li}. Thus one can believe that in the baryon
case, FSI may also be significant. In fact, our numerical results
show that without considering FSI, the branching ratio of
$\Lambda_c\rightarrow p\pi^0$ is only of order $10^{-6}$. However,
$\Lambda_c\rightarrow p\pi^0$ can be realized via
$\Lambda_c\rightarrow \Lambda K(K^*)\rightarrow p\pi^0$ or
$\Lambda_c\rightarrow n\pi^+(\rho^+) \rightarrow p\pi^0$ inelastic
final scattering,  and the resultant branching ratio could reach
order of $10^{-4}$. The results can be tested in the future
experiments.

Obviously our calculations depend on some phenomenological ansatz
and certain parameters, but all what we adopt in this work have
been widely used in literature and most of them have been tested
by experiments to some extent, hence they are somewhat trustworthy
or at least reasonable. Definitely, our results depend on the
chosen parameters, but not very sensitively. In other words we do
not fine-tune the numbers, therefore one can trust the order of
magnitude of the results.

The work is organized as follows. After the introduction, we
present our formulation and meanwhile very briefly review some
necessary knowledge. In Sec.III, we give our numerical results as
well as the adopted parameters and the last section is devoted to
our conclusion and discussion.\\

\noindent{\bf II. Formulation}\\

The weak effective hamiltonian is written as (for $c \rightarrow
q$ transition) \cite{Deshpande,Gilman}
\begin{equation}
H_{eff}={G_F\over\sqrt 2}\sum_{i=1}^6 V_{CKM}^ic_i(\mu)O_i,
\end{equation}
where $c_i\;(i=1,...,6)$ are the QCD modified Wilson coefficients
and the operators $O_i$ have the following expressions:
\begin{eqnarray}
O_1&=& \bar u_\alpha \gamma_\mu(1-\gamma_5)q_\beta\bar
q_\beta\gamma^\mu(1-\gamma_5)c_\alpha,\;\; O_2= \bar u
\gamma_\mu(1-\gamma_5)q\bar
q\gamma^\mu(1-\gamma_5)c,\nn\\
O_3&=& \bar u \gamma_\mu(1-\gamma_5)c \sum_{q'} \bar q'
\gamma^\mu(1-\gamma_5) q',\;\; O_4 = \bar u_\alpha
\gamma_\mu(1-\gamma_5)c_\beta \sum_{q'}
\bar q'_\beta \gamma^\mu(1-\gamma_5) q'_\alpha,\nn\\
O_5&=&\bar u \gamma_\mu(1-\gamma_5)c \sum_{q'} \bar q'
\gamma^\mu(1+\gamma_5)q',\;\; O_6 = \bar u_\alpha
\gamma_\mu(1-\gamma_5)c_\beta \sum_{q'}
\bar q'_\beta \gamma^\mu(1+\gamma_5) q'_\alpha,\nn\\
&& \label{add}
\end{eqnarray}
where $\alpha$ and $\beta$ are color indices, and $q'=u,\;d,\;s$.
In the Hamiltonian we have omitted the operators associated with
electroweak penguin diagrams.

The corresponding transition matrix elements are
\begin{equation}
<f|H_{eff}|i>= {G_F\over\sqrt 2}\sum_{i=1}^6 V_{CKM}^ic_i(\mu)
<f|O_i|i>.
\end{equation}
\\

In the limit $m_c \rightarrow \infty$, the baryonic transition
matrix element for $\Lambda_c\rightarrow f$ can be formed as
\begin{equation}
\label{mat} <f( p_f)|\bar
q\gamma_{\mu}(1-\gamma_5)c|\Lambda_c(v)>=
\overline{U}_f(p_f)[F_1(v\cdot p_f)+\rlap/{v} F_2(v\cdot p_f)]
\gamma_{\mu}(1-\gamma_5)U_{\Lambda_c}(v), \label{gg}
\end{equation}
where $v$ is the four-velocity of $\Lambda_c$, $q=u,d,s$, and
$U_f$ and $U_{\Lambda_c}$ are Dirac spinors for $f$ and
$\Lambda_c$, respectively. The two form factors $F_1$ and $F_2$
must be evaluated in concrete models.

To calculate the transition amplitude, it is usually assumed that
the process only occurs at the heavy quark arm while the light
flavor quark (diqaurk) stands as a spectator. Recent research
points out that the non-spectator contributions may be
important\cite{Bigi} for evaluating the lifetimes of $B$ and $D$
mesons. However, following literature, in the exclusive processes,
the spectator seems to be fully dominant and it is reasonable to
believe that the evaluation is sufficiently accurate if only the
spectator contribution is taken into account.\\

In the quark-diquark picture, the two-body baryonic wavefunction
can be written as \cite{Guo}
\begin{equation}
\psi_{i}(x_{1},{\bf k}_{\perp}) = N_{i}x_{1}x_{2}^3 {\rm exp}
[-b^{2} ({\bf k}_{\perp}^2+ m_{i}^{2}(x_{1}-x_{0i})^{2})].
\label{3b} \vspace{2mm}
\end{equation}
where $i=\Lambda_c, n, p$ or $\Lambda$, $x_1,x_2(x_2=1-x_1)$ are
the longitudinal momentum components of the quark and diqaurk,
${\bf k}_{\perp}$ is the transverse momentum of the constituents,
and $N_i$ is a normalization factor.\\

The two form factors $F_1$ and $F_2$ in eq.(\ref{mat}) are related
to an overlapping integral of the wavefunctions of $\Lambda_c$ and
the produced baryon \cite{Guo}:
\begin{eqnarray}
F_{1}&=& \frac{2E_{f}+m_{f}+m_{q}}{2(E_{f}+m_{q})}C_{s}I(\omega),
\non\\
F_{2}&=& \frac{m_{q}-m_{f}}{2(E_{f}+m_{q})}C_{s}I(\omega).
\label{3c} \vspace{2mm}
\end{eqnarray}
\noindent where $I(\omega)$ is the overlapping integral of the
wavefunctions of $\Lambda_c$ and $f$
\begin{eqnarray}
I(\omega) &=& \left(\frac{2}{\omega+1}\right)^{7/4}y^{-9/2} [A_f
K_6(\sqrt{2} b \epsilon)]^{-1/2} {\rm
exp}\left(-2b^2\epsilon^2\frac{\omega-1}
{\omega+1}\right)\times \non\\
&&\int_{-\frac{2b\epsilon}{\sqrt{\omega+1}}}^{y
-\frac{2b\epsilon}{\sqrt{\omega+1}}}{\rm d}z \;{\rm exp}(-z^2)
\left(y-\frac{2b\epsilon}{\sqrt{\omega+1}}-z\right)
\left(z+\frac{2b\epsilon}{\sqrt{\omega+1}}\right)^6, \label{3d}
\vspace{2mm}
\end{eqnarray}
where $y=bm_f\sqrt{\omega+1}$, $\omega$ is the velocity transfer
and $\omega=v\cdot p_f/m_f$.  $A_f$ and $K_6$ are defined as
\begin{eqnarray}
&& A_f=\int_{0}^{1}{\rm d}x\;x^6 (1-x)^2{\rm exp}[-2b^2 m_{f}^{2}
(x-\epsilon/m_f)^2],\non\\
&& K_6(\sqrt{2} b \epsilon)=\int_{-\sqrt{2} b \epsilon}^{\infty}
{\rm d }x \;{\exp}(-x^2)(x+\sqrt{2} b \epsilon)^6. \label{3e}
\vspace{2mm}
\end{eqnarray}

It is noted that here the limit $m_c\rightarrow \infty$ is taken.
In general, according to the Lorentz structure the transition
matrix element of $B_i({\bf p}_i)\rightarrow B_f({\bf p}_f)$ is
decomposed into the form
\begin{eqnarray}
\langle B_f(p_f)|V_\mu-A_\mu|B_i(p_i)\rangle &=& \bar{U}_f(p_f)
[f_1(q^2)\gamma_\mu+if_2(q^2)\sigma_{\mu\nu}q^\nu+f_3(q^2)q_\mu   \nonumber \\
&&
-(g_1(q^2)\gamma_\mu+ig_2(q^2)\sigma_{\mu\nu}q^\nu+g_3(q^2)q_\mu)\gamma_5]
U_i(p_i), \label{4a}
\end{eqnarray}
where $q=p_i-p_f$. Comparing eqs. (\ref{gg}) and (\ref{4a}) we
have relations
\begin{equation}
f_{1}(q^{2}) = g_{1}(q^{2}) = F_{1}(q^{2}) + \frac{m_{f}}
   {m_{i}}F_{2}(q^{2}) ,
\end{equation}
\begin{equation}
f_{2}(q^{2}) = g_{2}(q^{2}) = f_{3}(q^{2}) =
   g_{3}(q^{2}) = \frac{1}{m_{i}}F_{2}(q^{2}).
\end{equation}

Finally we reach the following formula:
\begin{eqnarray}
\langle B_f(p_f)|V_\mu-A_\mu|B_i(p_i)\rangle = \bar{U}_f
[a_1\gamma_\mu\gamma_5+a_2(p_i)_\mu\gamma_5
+b_1\gamma_\mu+b_2(p_i)_\mu]U_i,
\end{eqnarray}
where
\begin{eqnarray}
\non && a_1=-[F_1(q^2)+F_2(q^2)],\;\; a_2=-2F_2(q^2)/m_i;
\\  &&
\label{4b} b_1=[F_1(q^2)-F_2(q^2)],\;\; b_2=2F_2(q^2)/m_i.
\end{eqnarray}
\\

The process $\Lambda_c\rightarrow p\pi^0$ is Cabibbo-suppressed
and the reaction corresponds to the so-called inner emission which
is a color-suppressed process \cite{Stech}. In terms of the weak
effective hamiltonian we obtain
\begin{eqnarray}
\non {\cal M}(\Lamc \ra p\pi^0)&=& -{{G_F}\over \sqrt
{2}}V_{ub}V_{cb}^*
           {\Big [}(1+\frac{1}{N_c})(c_3+c_4)
           \lan \pi^0|\bar u\gamma^\mu(1-\gamma_5)u|0\ran
           \lan p|\bar u\gamma_\mu(1-\gamma_5)c|\Lamc \ran
       \\ \non  & &
           +(c_5+{c_6\over {N_c}})
           \lan \pi^0|\bar u\gamma^\mu(1+\gamma_5)u|0\ran
           \lan p|\bar u\gamma_\mu(1-\gamma_5)c|\Lamc \ran
       \\ \non   & &
          -2({c_5\over {N_c}}+c_6)
           \lan \pi^0|\bar u(1+\gamma_5)u|0\ran
           \lan p|\bar u(1-\gamma_5)c|\Lamc \ran{\Big ]}
        \\ \non   & &
       +{{G_F}\over \sqrt {2}}V_{ud}V_{cd}^*(c_1+{c_2\over {N_c}})
           \lan \pi^0|\bar d\gamma^\mu(1-\gamma_5)d|0\ran
           \lan p|\bar u\gamma_\mu(1-\gamma_5)c|\Lamc \ran
        \\ \non   & &
       -{{G_F}\over \sqrt {2}}V_{ub}V_{cb}^*{\Big [}(c_3+{c_4\over {N_c}})
          \lan \pi^0|\bar d\gamma^\mu(1-\gamma_5)d|0\ran
          \lan p|\bar u\gamma_\mu(1-\gamma_5)c|\Lamc \ran
        \\   & &
       +(c_5+{c_6\over {N_c}})
       \lan \pi^0|\bar d\gamma^\mu(1+\gamma_5)d|0\ran
       \lan p|\bar u\gamma_\mu(1-\gamma_5)c|\Lamc \ran{\Big ]},
\end{eqnarray}
where $N_c$ is the effective number of color which includes
non-factorizable effects.  After some simple manipulations, it is
recast as
\begin{eqnarray}
 {\cal M}(\Lamc \ra p\pi^0)&=&
    -{{G_F}\over \sqrt {2}}{\Big [} V_{ud}V_{cd}^*(c_1+{c_2\over {N_c}})
   +V_{ub}V_{cb}^*({c_3\over {N_c}}+c_4)
   {\Big ]}\;
   {{if_\pi}\over {\sqrt{2}}}
  \bar{U}_p (
   A^{\prime}+B^{\prime}\gamma_5
    )U_{\Lamc} \nonumber  \\  & &
 +\sqrt {2}G_F V_{ub}V_{cb}^*({c_5\over {N_c}}+c_6)
  (-\frac{if_{\pi}}{2\sqrt{2}})\;
 \bar {U_p}(A^{\prime\prime}
 +B^{\prime\prime}\gamma_5)U_{\Lamc}
 \non  \\ &=&
 i\; \bar {U_p}(A+B\gamma_5)U_{\Lamc},
\end{eqnarray}
where
\begin{eqnarray}
\non & & A=-{{G_F}\over 2}f_\pi {\Big \{}
  {\Big [} V_{ud}V_{cd}^*(c_1+{c_2\over {N_c}})
   +V_{ub}V_{cb}^*({c_3\over {N_c}}+c_4)
   {\Big ]}A^{\prime}\;
  +V_{ub}V_{cb}^*({c_5\over {N_c}}+c_6)\;
  A^{\prime\prime} {\Big \}},
\\    \non   & &
B=-{{G_F}\over 2}f_\pi {\Big \{}
   {\Big [} V_{ud}V_{cd}^*(c_1+{c_2\over {N_c}})
   +V_{ub}V_{cb}^*({c_3\over {N_c}}+c_4)
   {\Big ]}B^{\prime}\;
   +V_{ub}V_{cb}^*({c_5\over {N_c}}+c_6)\;
  B^{\prime\prime}{\Big \}},
\\   \non  & &
  A^{\prime\prime}=\frac{m^2_{\pi}}{m_u(m_c-m_u)}A^{\prime}\;,
  B^{\prime\prime}=-\frac{m^2_{\pi}}{m_u(m_c+m_u)}B^{\prime},
\\   \non  & &
  A^{\prime}=b_1(m_{\Lamc}-m_p)+b_2\frac{m^2_{\Lamc}+m^2_{\pi}-m^2_p}{2},
  B^{\prime}=-a_1(m_{\Lamc}+m_p)+a_2\frac{m^2_{\Lamc}+m^2_{\pi}-m^2_p}{2},
\\   \non  & &
a_1=-[F_1(m^2_{\pi})+F_2(m^2_{\pi})],\;\;
a_2=-2F_2(m^2_{\pi})/m_{\Lamc},
\\  &&
b_1=[F_1(m^2_{\pi})-F_2(m^2_{\pi})],\;\;
b_2=2F_2(m^2_{\pi})/m_{\Lamc}. \label{ggg}
\end{eqnarray}
\\

The expressions for $\Lambda_c\rightarrow n\pi^+(\rho^+)$ are
similar to that for $\Lambda_c\rightarrow p\pi^0$, so we omit the
details and show the result directly,
\begin{eqnarray}
      {\cal M}(\Lamc \ra n\pi^+)
&=&
    i\; \bar{U}_n (
     A+B\gamma_5
       )U_{\Lamc},
\end{eqnarray}
here $A, B$ take values as follows
\begin{eqnarray}
\non A&=&{{G_F}\over \sqrt {2}}
     [V_{ud}V_{cd}^*({c_1\over {N_c}}+c_2)
      -V_{ub}V_{cb}^*(\frac{c_3}{N_c}+c_4)]
     {f_\pi} A^{\prime}
\\ \non & &
       -V_{ub}V_{cb}^*({c_5\over {N_c}}+c_6)
      \frac{\sqrt {2}G_Ff_{\pi}m^2_{\pi}}{(m_u+m_d)(m_c-m_d)}
     A^{\prime},
       \\    \non
B&=&{{G_F}\over \sqrt {2}}
     [V_{ud}V_{cd}^*({c_1\over {N_c}}+c_2)
      -V_{ub}V_{cb}^*(\frac{c_3}{N_c}+c_4)]
     {f_\pi} B^{\prime}
\\  & &
       +V_{ub}V_{cb}^*({c_5\over {N_c}}+c_6)
      \frac{\sqrt {2}G_Ff_{\pi}m^2_{\pi}}{(m_u+m_d)(m_c+m_d)}
     B^{\prime},
\label{ab1}
\end{eqnarray}
where $A'$ and $B'$ have the same forms as those in eq.(\ref{ggg})
with $m_p$ being replaced by $m_n$. In the same way, we have
\begin{eqnarray}
\non
&& {\cal M}(\Lamc \ra n\rho^+) = \nonumber\\
&& {{G_F}\over \sqrt {2}}
        [V_{ud}V_{cd}^*({c_1\over {N_c}}+c_2)
         -V_{ub}V_{cb}^*(\frac{c_3}{N_c}+c_4)]
         \lan \rho^+|\bar u\gamma^\mu d|0\ran
         \lan n|\bar d\gamma_\mu(1-\gamma_5)c|\Lamc \ran.
\end{eqnarray}
Finally,
\begin{eqnarray}
{\cal M}(\Lamc \ra n\rho^+)&=& \bar{U}_n \epsilon^{*\mu}_\rho
{\Big [}A_1\gamma_\mu\gamma_5+A_2(p_{\Lamc})_\mu\gamma_5
+B_1\gamma_\mu+B_2(p_{\Lamc})_\mu{\Big ]}U_{\Lamc}, \label{gggg}
\end{eqnarray}
where $A_i=a_i \eta$,  $B_i=b_i \eta$ (i=1, 2), and
\begin{equation}
\eta=  {{G_F}\over \sqrt {2}}
       [V_{ud}V_{cd}^*({c_1\over {N_c}}+c_2)
        -V_{ub}V_{cb}^*(\frac{c_3}{N_c}+c_4)] f_\rho m_\rho.
\end{equation}
\\

For $\Lambda_c\rightarrow \Lambda K^+(K^{+*})$, the formulas are
in analog to those given above. We have
\begin{eqnarray}
\non & & {\cal M}(\Lamc \ra \Lam K^+)
\nonumber \\
&=&
        {{G_F}\over \sqrt {2}}
        [V_{us}V_{cs}^*({c_1\over {N_c}}+c_2)
         -V_{ub}V_{cb}^*(\frac{c_3}{N_c}+c_4)]
        {if_K}\bar{U}_\Lam (
        A^{\prime}+B^{\prime}\gamma_5
         )U_{\Lamc}
 \nonumber \\  & &
        -V_{ub}V_{cb}^*({c_5\over {N_c}}+c_6)
      \frac{i\sqrt {2}G_Ff_{K}m^2_{K}}{(m_u+m_s)(m_c^2-m_s^2)}
          \bar{U}_\Lambda {\Big [}
     (m_c+m_s)A^{\prime}
     -(m_c-m_s)B^{\prime}\gamma_5
    {\Big ]}U_{\Lamc} \nonumber
   \\  &=&
     i\;\bar{U}_\Lam (
     A+B\gamma_5
       )U_{\Lamc},
\end{eqnarray}
where $A, B$ are given by
\begin{eqnarray}\label{ab2}
A&=&{{G_F}\over \sqrt {2}}
     [V_{us}V_{cs}^*({c_1\over {N_c}}+c_2)
      -V_{ub}V_{cb}^*(\frac{C_3}{N_c}+c_4)]
     {f_K} A^{\prime},
\nonumber \\
 & &
       -V_{ub}V_{cb}^*({c_5\over {N_c}}+c_6)
      \frac{\sqrt {2}G_Ff_{K}m^2_{K}}{(m_u+m_s)(m_c-m_s)}
     A^{\prime},
       \nonumber \\
B&=&{{G_F}\over \sqrt {2}}
     [V_{us}V_{cs}^*({c_1\over {N_c}}+c_2)
      -V_{ub}V_{cb}^*(\frac{C_3}{N_c}+c_4)]
     {f_K} B^{\prime}
\nonumber \\  & &
     +V_{ub}V_{cb}^*({c_5\over {N_c}}+c_6)
      \frac{\sqrt {2}G_Ff_{K}m^2_{K}}{(m_u+m_s)(m_c+m_s)}
     B^{\prime},
\end{eqnarray}
with
\begin{eqnarray}
 &&
A^{\prime}=b_1(m_{\Lamc}-m_\Lam)+b_2(p_{K}\cdot p_{\Lamc})
          =b_1(m_{\Lamc}-m_\Lam)+b_2\frac{m^2_{\Lamc}+m^2_{K}-m^2_\Lam},
\nonumber \\  && B^{\prime}=a_1(-m_{\Lamc}-m_\Lam)+a_2(p_{K}\cdot
p_{\Lamc})
          =-a_1(m_{\Lamc}+m_\Lam)+a_2\frac{m^2_{\Lamc}+m^2_{K}-m^2_\Lam}{2}.
\end{eqnarray}

For   $\Lambda_c\rightarrow \Lambda K^{*+}$, we have
\begin{eqnarray}
{\cal M}(\Lamc \ra \Lam K^{*+})&=& \bar{U}_\Lam
\epsilon^{*\mu}_{K^*} {\Big
[}A_1\gamma_\mu\gamma_5+A_2(p_{\Lamc})_\mu\gamma_5
+B_1\gamma_\mu+B_2(p_{\Lamc})_\mu{\Big ]}U_{\Lamc},
\end{eqnarray}
where $A_i=a_i \tilde{\eta}$ and $B_i=b_i \tilde{\eta} $ (i=1, 2),
and
\begin{equation}
\tilde{\eta}=  {{G_F}\over \sqrt {2}}
       [V_{us}V_{cs}^*({c_1\over {N_c}}+c_2)
        -V_{ub}V_{cb}^*(\frac{c_3}{N_c}+c_4)] f_{K^*}
        m_{K^*}.
\end{equation}

With all the amplitudes, we can immediately obtain the decay
widths and corresponding branching ratios if there were no final
state interaction. But in fact, the FSI effects can drastically
change the results. In the following, we will present the
formulas for the FSI effects.\\

Since the final state interactions may change the whole picture
for some processes, one needs to take into account these effects
seriously. Unfortunately, the FSI processes are governed by
non-perturbative QCD, so we cannot simply start from any
underlying theory to carry out the computations yet. One could
extract valuable information from data. Theoretically, we may also
use some phenomenological approaches to study the order of
magnitude of the FSI effects.

There are several ways to evaluate the FSI. A usually adopted
method is the one-particle exchange model \cite{Ani} and another
one is that in which the produced primary particles of the decay
scatter into the final products via exchanging the Regge poles
\cite{Collins}. The first method is straightforward and simple,
and we employ it in this work.

In this method, the effective vertices can be obtained from the
chiral Lagrangian \cite{Georgi1}. The coefficients in the chiral
Lagrangian are obtained directly from data where all external
particles are on their mass shells. Instead, for our case the
exchanged mesons or baryons are obviously off mass shell. To
compensate the off-shell effects, one needs to introduce a form
factor at the effective vertices \cite{Gort}.

The Feynman diagrams for the FSI are shown in Fig.1.

\begin{center}
\begin{figure}
\vspace*{2.6cm}
\begin{picture}(400,200)(0,-60)
\put(28,200){\line(1,0){30}} \put(62,200){\circle{8}}
\put(66,200){\line(3,2) {45}} \put(66,200){\line(3,-2){45}}
\put(115,168){\circle{8}} \put(115,232){\circle{8}}
\put(115,172){\line(0,1){56}} \put(119,168){\line(1,0){38}}
\put(119,232){\line(1,0){38}}

\put(85,186.8){\vector(3,-2){2}} \put(85,212.5){\vector(3,2){2}}

\put(10,198){$\Lamc$} \put(77,168){$\pi^+$} \put(77,220){$n$}
\put(118,195){$\rho^+$} \put(158,228){$p$} \put(158,163){$\pi^0$}
\put(108,140){(a)}

\put(228,200){\line(1,0){30}} \put(262,200){\circle{8}}
\put(266,200){\line(3,2) {45}} \put(266,200){\line(3,-2){45}}
\put(315,168){\circle{8}} \put(315,232){\circle{8}}
\put(315,172){\line(0,1){56}} \put(319,168){\line(1,0){38}}
\put(319,232){\line(1,0){38}} \put(285,186.8){\vector(3,-2){2}}
\put(285,212.5){\vector(3,2){2}}

\put(210,198){$\Lamc$} \put(272,168){$\rho^{+}$}
\put(272,220){$n$} \put(318,195){$\pi^-$} \put(358,228){$p$}
\put(358,163){$\pi^0$} \put(320,140){(b)}

\put(28,60){\line(1,0){30}} \put(62,60){\circle{8}}
\put(66,60){\line(3,2) {45}} \put(66,60){\line(3,-2){45}}
\put(115,28){\circle{8}} \put(115,92){\circle{8}}
\put(115,32){\line(0,1){56}} \put(119,28){\line(1,0){38}}
\put(119,92){\line(1,0){38}} \put(85,46.8){\vector(3,-2){2}}
\put(85,72.5){\vector(3,2){2}} \put(10,58){$\Lamc$}
\put(77,28){$K^{+}$} \put(77,80){$\Lam$} \put(118,55){$K^{*-}$}
\put(158,88){$p$} \put(158,23){$\pi^0$} \put(108,-10){(c)}
\put(228,60){\line(1,0){30}} \put(262,60){\circle{8}}
\put(266,60){\line(3,2) {45}} \put(266,60){\line(3,-2){45}}
\put(315,28){\circle{8}} \put(315,92){\circle{8}}
\put(315,32){\line(0,1){56}} \put(319,28){\line(1,0){38}}
\put(319,92){\line(1,0){38}} \put(285,46.8){\vector(3,-2){2}}
\put(285,72.5){\vector(3,2){2}} \put(210,58){$\Lamc$}
\put(277,28){$K^{*+}$} \put(277,80){$\Lam$} \put(318,55){$K^{-}$}
\put(358,88){$p$} \put(358,23){$\pi^0$} \put(320,-10){(d)}

\put(85,-50){Fig.1  The Final State Interaction for $\Lam_c \to p
\pi^0 $}
\end{picture}

\end{figure}
\end{center}


In fact, there are some u-channel diagrams, namely in Fig.1,  $p$
and  $\pi^0$ may exchange positions and the intermediate bosons
$\rho(K^*)$ or $\pi(K)$ are replaced by baryons $p$ or $\Lambda$.
But a direct calculation indicates that the u-channel
contributions are only at most  $10^{-3}\sim 10^{-2}$ of the
t-channel, so we omit the u-channel part in later calculations for
simplicity.

Moreover, the direct decay of $\Lambda_c\rightarrow p\pi^0$ is
color suppressed compared to $\Lambda_c\rightarrow n\pi^+$ etc.,
and hence has smaller amplitudes. Thus we only consider
$\Lambda_c\rightarrow n\pi^+,n\rho^+,\Lambda K^+,\Lambda K^{+*}
\rightarrow p\pi^0$, but not the opposite directions.\\

For $\Lambda_c\rightarrow p\pi^0$, the amplitude with final state
interactions can be written as
\begin{eqnarray}
A^{FSI}_{n\pi^+ \to p \pi^0} &= & \frac{1}{2}\int \frac{d^3
p_1}{(2\pi)^3 2E_1}\frac{d^3 p_2} {(2\pi)^3 2E_2}(2\pi)^4
\delta^4(p_1+p_2-p_{\Lamc})
\nonumber \\
 & & \cdot A(\Lamc \to n\pi^+)
  \lan n\pi^+ |S|p \pi^0 \ran,
\end{eqnarray}
where for clarity, we set $p_1=p_n, p_2=p_{\pi^+}, p_3=p_p$ and
$p_4=p_{\pi^0}$. In terms of the Cutkosky cutting rule\cite{Cut},
we can easily calculate the inelastic scattering amplitude.

The effective $\rho NN$ has the following form \cite{Zhu}:
\begin{equation}
\label{6aa} {\cal L}_{\rho NN}=g_{\rho NN}
\epsilon_{\rho}^{\mu}{\bar N}
    {\Big (}
    \gamma_\mu+\kappa_{\rho}\frac{i\sigma_{\mu \nu}q^\nu}{2m_N}
    {\Big )}N.
\end{equation}
With this vertex, we have
\begin{eqnarray}
\label{6a} A^{FSI}_{n\pi^+ \to p \pi^0} &= &
    \frac{1}{2}\int \frac{d^4 p_1}{(2\pi)^2}
     \delta(p_1^2-m^2) \delta(p_1^2-m^2)
     \frac{iF(k^2)g_{\rho NN}g_{\rho\pi\pi}}{k^2-m_{\rho}^2}
     {\bar U}_p(p_3)\epsilon^{\mu}_{\rho}
    {\Big (}
     \gamma_\mu+\kappa_{\rho}\frac{i\sigma_{\mu \nu}q^\nu}{2m_N}
    {\Big )}
\nonumber \\  & &
    \cdot \epsilon^{*\alpha }_{\rho}
    {(p_2+p_4)}_{\alpha}(\rlap/{p_1}+m_1)
    i(A+B\gamma_5)U_{\Lamc} (p_{\Lamc}).
\end{eqnarray}
Here $F(k^2)$ is a form factor\cite{Zou,Dai}. In fact, the
effective couplings at the vertices are obtained from real
processes where all particles are on mass shell, but in our case,
at least one of the particles (the exchanged $\rho$ meson) is off
shell. Therefore, we introduce this form factor as a compensation
to the off shell effects. Following the literature, in our
calculation we take
$$F(k^2)=\left[
(\alpha^2-m_{\rho}^2)/(\alpha^2-p_{\rho}^2)\right]^2,$$ where
$\alpha$ is a parameter.

The explicit expressions for $A^{FSI}_{n\pi^+ \to p \pi^0}$ as
well as those for other inelastic final state interactions,
$\Lambda_c\rightarrow n\rho^+\rightarrow p\pi^0$,
$\Lambda_c\rightarrow \Lambda K^+\rightarrow p\pi^0$ and
$\Lambda_c\rightarrow \Lambda K^{*+}\rightarrow p\pi^0$, are
rather complicated, so we collect them in the appendix.
\\

\noindent{\bf III. The numerical results}\\

As inputs, we take the $c_i$'s in the weak effective Lagrangian
from literature \cite{Deshpande,Flei} where the renormalization is
carried out to one-loop level:
$$c_i(m_c)<O_i(m_c)>=c_i'<O_i>^{tree},$$
where $c_i'$ are effective Wilson coefficients.

In the wavefunction $\psi_i(x_1,{\bf k}_{\perp})$, we take
$b=1.77$ GeV and 1.18 GeV corresponding to $<{\bf
k}^2_{\perp}>^{1/2}=400$ MeV and 600 MeV, respectively. $x_{0i}
(x_{0i}=1-m_D/m_i$, $i=\Lambda_c, n$ or $\Lambda$) is calculated
with the diquark mass $m_D$ being taken the value of 600 MeV. We
let the parameter $\alpha$ in the form factor $F(q^2,\alpha)$ vary
in a reasonable region. The results are not very sensitive to
these parameters as shown below.

We also set $m_u\sim m_d\sim 10 $ MeV, $m_s=150$ MeV, $m_c=1.35$
GeV. For the hadron decay constants, $f_{\pi}=132$ MeV, $f_K=160$
MeV, $f_{\rho}=210$ MeV, and $f_{K^*}=221$ MeV.

The CKM entries are  in the Wolfenstein parametrization and
all values are taken from the Particle Data book.\\

a. Without the final state interaction.

We first present the numerical results when the FSI effects are
turned off. The numerical results are shown in Table.1 and
Table.2.\\

\begin{center}
\vspace*{0.5cm}
\begin{tabular}{|c|c|c|c|c|c|c|}
\hline \hline \multicolumn{2}{|c|}{Products}
              & $\;\;\; p\pi^0\;\;\;$   & $\;\;\; n\pi^+\;\;\;$
              & $\;\;\; n\rho^+\;\;\;$  & $\;\;\; \Lam K^+\;\;\;$
              & $\;\;\; \Lam K^{*+}\;\;\;$
\\
\hline \hline $b=1.77$     & $\;\;\;F_1\;\;\;$
             &  0.207      & 0.209
             &  0.264      & 0.301
             &  0.384
\\
\cline{2-7}
   GeV       & $\;\;\;F_2\;\;\;$
             & -0.0533     & -0.0538
             & -0.0730     & -0.0728
             & -0.0989
\\
\hline $b=1.18$      & $\;\;\;F_1\;\;\;$
              &  0.145     & 0.146
              &  0.165     & 0.208
              &  0.236
\\
\cline{2-7}
   GeV        & $\;\;\;F_2\;\;\;$
              & -0.0372    & -0.0375
              & -0.0456    & -0.0503
              & -0.0609
\\
\hline
\end{tabular}
\\
\vspace*{0.5cm} {\small Table 1. The form factors $F_1$ and $F_2$
as $b$ takes two values.}
\end{center}

\begin{center}
\vspace*{0.5cm}
\begin{tabular}{|c|c|c|c|c|c|c|}
\hline \hline
             Branching ratos
             & $\;\;\; p\pi^0 \;\;\;$
             & $\;\;\; n\pi^+ \;\;\;$
             & $\;\;\; n\rho^+ \;\;\;$
             & $\;\;\; \Lam K^+ \;\;\;$
             & $\;\;\; \Lam K^{*+} \;\;\;$
\\
\hline \hline
             $b=1.77$ GeV
             &  $4.35\times 10^{-6}$
             &  $2.10\times 10^{-4}$
             &  $3.57\times 10^{-4}$
             &  $3.86\times 10^{-4}$
             &  $4.91\times 10^{-4}$
\\
\hline
              $b=1.18$ GeV
              &  $2.14\times 10^{-6}$
              &  $1.03\times 10^{-4}$
              &  $1.39\times 10^{-4}$
              &  $1.84\times 10^{-4}$
              &  $1.85\times 10^{-4}$
\\
\hline
\end{tabular}
\\
\vspace*{0.5cm} {\small Table 2. The branching ratios. Here we
have $\tau_{\Lamc}=0.206 $ ps \cite{PDG}.}
\end{center}

\vspace{0.8cm}

b. Taking the FSI effects into account.

The coupling constants of the strong interaction at the vertices
take the values $g_{K^*K\pi}=5.8$, $g_{\rho\pi\pi}=6.1$
\cite{Zou,Du}, $g_{\rho NN}=2.5, \kappa_{\rho}=8.0$ \cite{Zhu},
$g_{\pi NN}=13.6$ \cite{Zhu1}, $g_{K\Lam N}=-10$ \cite{Aliev}. To
determine the coupling $g_{K^{*}\Lam N}$, we assume the equality
$g_{K^{*}\Lam N}/g_{\rho NN}=g_{K\Lam N}/g_{\pi NN}$. In the
compensation form factors $F(\alpha)$, the parameter $\alpha$
takes values in the reasonable region\cite{Gort,Zou,Du} as
$1.2-2.0$ GeV. In our concrete calculations, we employ
$\alpha=0.9$ GeV, $1.0$ GeV, and $1.2$ GeV, respectively. From
\cite{Aliev} we have $|g_{K\Lam N}/g_{K\Sigma N}|=12$, thus we do
not need to consider the contribution of $\Lamc \to \Sigma K$ to
$\Lamc \to p\pi$ via final state interaction. The calculations are
done for the case of maximum interference.

The numerical results with FSI effects are listed in table 3.   \\

\begin{center}
\vspace*{0.5cm}
\begin{tabular}{|c|c|c|c|c|}
\hline\hline
 & without  &\multicolumn{3}{|c|}{with FSI effects}
\\  \cline{3-5}
&  FSI effects
     & $\alpha=0.9$ GeV
   & $\alpha=1.0$ GeV
   & $\alpha=1.2$ GeV
\\  \hline \hline
      $b=1.77$ GeV
   &$4.35\times 10^{-6}$
   &$3.14\times 10^{-4}$
   &$2.92\times 10^{-4}$
   &$3.55\times 10^{-4}$
\\  \hline
     $b=1.18$ GeV
   &$2.14\times 10^{-6}$
   &$1.20\times 10^{-4}$
   &$1.11\times 10^{-4}$
   &$1.50\times 10^{-4}$
\\ \hline
\end{tabular}
\vspace*{0.5cm}
\\
{\small Table 3. The branching ratio of $\Lamc \to p\pi^0$ with
and without the FSI effects. }
\end{center}

\vspace{0.6cm}

\noindent{IV. Conclusion and discussion}\\

It is widely recognized that non-perturbative QCD is, so far, an
unsolved problem, but it is entangled with the effects of the
fundamental mechanisms. We need to explore its properties in the
phenomenological studies and gradually get better understanding of
it.

The processes where only mesons are involved have been carefully
investigated, including the final state interactions, even though
there is still no way to evaluate them exactly from underlying
theory. On the other hand, for the baryon case, the situation is
more complicated because there are three valence quarks in a
baryon.

It is reasonable to consider the diquark structure in baryons,
which may be one of the possible configurations of three quarks.
For a heavy baryon with one heavy quark and two light quarks, the
two light flavors constitute a color $\bar 3$ diquark where the
interaction is attractive, and the heavy one acts as a color field
source. This configuration may be expected as the dominant one.
Thus in this work we employ this physical picture to carry out the
calculations. Definitely, this configuration does not possess
100\% probability, so we might introduce a phenomenological
parameter which is less than unity, but close to, for describing
the deviation. Therefore, the branching ratios we obtained in the
text must be multiplied by this parameter. However, the relative
ratios of $\Lambda_c\rightarrow p\pi^0$ to  $\Lambda_c\rightarrow
n\pi^+$ does not depend on it and this fact can be tested in
experiments. As a matter of fact, some authors also used the
diqaurk picture to evaluate the decays of heavy flavors to lighter
ones recently \cite{Hou}.

Moreover, it is easy to accept that in the spectator picture, the
heavy quark turns into a lighter one, there could be a possibility
that a light quark in the light diquark breaks out and combines
with the newly produced quark to constitute another diquark. But
from the physics intuition, this probability should be smaller
than retaining the original diquark structure as spectating the
heavy quark turn into a light one, so we assume that the light
diquark serves as a spectator and does not change after the weak
transition.

When we calculate weak transitions, we use the factorization
ansatz, which is not accurate. Therefore the absolute values of
the obtained branching ratios may decline from the real values by
a certain amount. But according to the information we have learned
in the meson case, for such exclusive processes, factorization
works. We will pursue this problem in our later works.

In this work, we show that the FSI effects are important, not only
for the meson cases, but also for baryon-involved processes. It is
indicated that without considering the FSI, the estimated
branching ratio of $\Lambda_c\rightarrow p\pi^0$ is almost two
orders smaller than that of $\Lambda_c\rightarrow n\pi^+$, but
with FSI, they could have the same order. The numerical results
show that the theoretically calculated values depend on the
parameters $\alpha$ and $b$ in the model employed here. This
originates from the lack of knowledge of non-perturbative QCD, and
we cannot fix them at present. However, as long as they fall in a
reasonable region, the order of magnitude does not change.

For evaluating the FSI effects, we only consider the absorptive
part of the triangle diagram. It is the real final state process
because both hadron which are produced via the weak transition are
on their mass shell. But definitely, the dispersive part of the
triangle which needs to be renormalized would also bring up some
effects and we ignore them in this calculation. Higher order FSI
may also play roles and it can induce a strong phase and cause CP
violations. We have investigated the possible CP violation in the
meson case \cite{Zou,Dai} because there are data about the
scattering amplitude and phases. In this concerned case, it is
impossible to carry out accurate calculation yet, so we can only
trust it to the order of magnitude.

By including the final state interaction, we show that the
branching ratios of $\Lambda_c\rightarrow p\pi^0$ could be greatly
enhanced by almost two orders and it is similar to the meson case,
as $D^0\rightarrow K^0\bar K^0$ is much suppressed compared to
$D^0\rightarrow K^+ K^-$, but measurements indicate that they are
at the same order. This is obviously due to the FSI effects.
Therefore when we evaluate the processes for baryons, we cannot
ignore  the FSI, as we understand in the meson case.

As pointed out above, many uncertainties exist in the theoretical
calculations, so for a better understanding, we need more data as
well as a more reliable theory for the non-perturbative QCD.
Fortunately, the BEPC, CLEO and other high energy accelerators
such as the B-factory and even TEVATRON and LHC will provide much
more data in the future and we will be able to use them to
determine the parameters involved in the phenomenological models
and surely test the model itself. \\

\noindent{Acknowledgement:}

This work is partially supported by the National Natural Science
Foundation of China and the Australian Research Council.\\

\vspace{1cm}

\noindent{\bf Appendix}

In this appendix we present some details of the formulation about
the final state inelastic scattering amplitudes.

(1) For $\Lambda_c\rightarrow n\pi^+\rightarrow p\pi^0$,
\begin{eqnarray}
A^{FSI}_{n\pi^+ \to p \pi^0}
        &= &  \non
     \frac{1}{2}\int \frac{d^4 p_1}{(2\pi)^2}
     \delta(p_1^2-m^2) \delta(p_1^2-m^2)
     \frac{iF(k^2)g_{\rho NN}g_{\rho\pi\pi}}{k^2-m_{\rho}^2}
\\  \non
& &  \cdot
    {\bar U}_p(p_3)
    {\Big \{}
    {\Big [}
    -4p_{\Lamc}\cdot p_1+m_1^2
    -\frac{\kappa_{\rho}}{m_N}
    (m_1^2\rlap/p_3-m_1^2\rlap/p_{\Lamc}-2p_{\Lamc}\cdot p_1\rlap/p_3)
    -\frac{m_1^2(m_2^2-m_4^2)}{m^2_{\rho}}
    {\Big ]}
\\  \non
& &
    +m_1
    {\Big [}
    -(2\rlap/p_{\Lamc}-\rlap/p_3)
    +\frac{\kappa_{\rho}}{m_N}(p_2\cdot p_4-M^2+2p_{\Lamc}\cdot p_1
    +\rlap/p_3\rlap/p_{\Lamc})
\\  \non
& &
    +\frac{(m_2^2-m_4^2)}{m^2_{\rho}}\rlap/p_3
    {\Big ]}
    {\Big \}}
    i(A+B\gamma_5)U_{\Lamc} (p_{\Lamc})
\\   \non
&= &
    \frac{1}{2}\int \frac{d^4 p_1}{(2\pi)^2}
     \delta(p_1^2-m^2) \delta(p_1^2-m^2)
     \frac{iF(k^2)g_{\rho NN}g_{\rho\pi\pi}}{k^2-m_{\rho}^2}
\\  \non
& &  \cdot
    {\bar U}_p(p_3)
    i(A H_1^a+B H_2^a \gamma_5)U_{\Lamc} (p_{\Lamc});
\\
&= &
    \int^1_{-1}
     \frac{d(cos\theta)|\overrightarrow p_1|}{16\pi M}
     \frac{iF(k^2)g_{\rho NN}g_{\rho\pi\pi}}{k^2-m_{\rho}^2}
     {\bar U}_p(p_3)
    i(A H_1^a+B H_2^a \gamma_5)U_{\Lamc} (p_{\Lamc}),
\end{eqnarray}
where $\theta$ is the angle spanned between ${\bf p}_1$ and ${\bf
p}_3$, $k^2=(p_1-p_3)^2=m_1^2+m_3^2-2E_1E_3 +2|\overrightarrow
p_1||\overrightarrow p_3|\cos(\theta)$, and $|\overrightarrow
p_1|=\lambda^{1/2}(M^2,m_1^2,m_2^2)/(2M)$.
$\lambda(x,y,z)=x^2+y^2+z^2-2xy-2yz-2xz$ is the well-known
function.In the above expression, $A, B$ take the values in
Eq.(\ref{ab1}) and $H_1^a$, $H_2^a$ are
\begin{eqnarray}
H_1^a &= &  \non
    {\Big [}
    -4p_{\Lamc}\cdot p_1+m_1^2
    -\frac{\kappa_{\rho}}{m_N}
    (m_1^2m_3-m_1^2M-2p_{\Lamc}\cdot p_1 m_3)
    -\frac{m_1^2(m_2^2-m_4^2)}{m^2_{\rho}}
    {\Big ]}
\\  \non
& &
    +m_1
    {\Big [}
    -2M+m_3
    +\frac{\kappa_{\rho}}{m_N}(p_2\cdot p_4-M^2+2p_{\Lamc}\cdot p_1
    +m_3M)
    +\frac{(m_2^2-m_4^2)m_3}{m^2_{\rho}}
    {\Big ]},
\\    \non
H_2^a &= &
    {\Big [}
    -4p_{\Lamc}\cdot p_1+m_1^2
    -\frac{\kappa_{\rho}}{m_N}
    (m_1^2m_3+m_1^2M-2p_{\Lamc}\cdot p_1 m_3)
    -\frac{m_1^2(m_2^2-m_4^2)}{m^2_{\rho}}
    {\Big ]}
\\  \non
& &
    +m_1
    {\Big [}
    2M+m_3
    +\frac{\kappa_{\rho}}{m_N}(p_2\cdot p_4-M^2+2p_{\Lamc}\cdot p_1
    -m_3M)
    +\frac{(m_2^2-m_4^2)m_3}{m^2_{\rho}}
    {\Big ]}.
\end{eqnarray}

(2) For $\Lambda_c\rightarrow n\rho^+\rightarrow p\pi^0$ we have
similar expressions
\begin{eqnarray}
\label{7a} \non A^{FSI}_{n\rho^+ \to p \pi^0} &= &
   \frac{1}{2}\int \frac{d^3 p_1}{(2\pi)^3 2E_1}\frac{d^3 p_2}
    {(2\pi)^3 2E_2}(2\pi)^4 \delta^4(p_1+p_2-p_{\Lamc})
     \cdot A(\Lamc \to n\pi^+)
     \lan n\rho^+ |S|p \pi^0 \ran
\\ \non
&= &
    \frac{1}{2}\int \frac{d^4 p_1}{(2\pi)^2}
     \delta(p_1^2-m^2) \delta(p_1^2-m^2)
     \frac{iF(k^2)g_{\pi NN}g_{\rho \pi \pi}}{k^2-m_{\pi}^2}
     i{\bar U}_p(p_3)\gamma_5
     (\rlap/p_1+m_1)
\\
& &
    \cdot \epsilon^{\mu }_{\rho}
    {(k+p_4)}_{\mu}
      \epsilon^{*\nu}_\rho
    {\Big [}
    A_1\gamma_\nu\gamma_5+A_2(p_1)_\nu\gamma_5
    +B_1\gamma_{\nu}+B_2(p_1)_\nu
    {\Big ]}
    U_{\Lamc},
\end{eqnarray}
where $p_1=p_n, p_2=p_{\rho^+}, p_3=p_p$ and $p_4=p_{\pi^0}$.
Eq.(\ref{7a}) is recast as
\begin{eqnarray}
\non
    A^{FSI}_{n\rho^+ \to p \pi^0}
&= &
    \frac{1}{2}\int \frac{d^4 p_1}{(2\pi)^2}
    \delta(p_1^2-m^2) \delta(p_1^2-m^2)
    \frac{iF(k^2)g_{\pi NN}g_{\rho \pi \pi}}{k^2-m_{\pi}^2}
    i{\bar U}_p(p_3)
    [A_b+B_b\gamma_5]
    U_{\Lamc}
\\
&= &
   \int^1_{-1} \frac{d(cos\theta)|\overrightarrow p_1|}{16\pi M}
   \frac{iF(k^2)g_{\pi NN}g_{\rho \pi \pi}}{k^2-m_{\pi}^2}
   i{\bar U}_p(p_3)
   [A_b+B_b\gamma_5]
   U_{\Lamc},
\end{eqnarray}
and $A_b, B_b$ are defined as
\begin{eqnarray}
\non A_b &= &
    2m_1
    {\Big \{}
     A_1(M-m_3)
    +A_2p_1\cdot (p_3-p_{\Lamc})
    +\frac{p_2\cdot p_4}{m_2^2}
    {\Big [}
   -A_1 M
   +A_2p_1\cdot (p_{\Lamc}-p_1)
    {\Big ]}
    {\Big \}}
\\   \non
& &
    +2
    {\Big \{}
    2p_1\cdot p_3A_1
    -\frac{p_2\cdot p_4}{m_2^2}
    m_1^2A_1
    {\Big \}},
\\  \non B_b
&= &
    2m_1
    {\Big \{}
     B_1(-M-m_3)
    +B_2p_1\cdot (p_3-p_{\Lamc})
    +\frac{p_2\cdot p_4}{m_2^2}
    {\Big [}
    B_1 M
   +B_2p_1\cdot (p_{\Lamc}-p_1)
    {\Big ]}
    {\Big \}}
\\   \non
& &
    +2
    {\Big \{}
     2p_1\cdot p_3B_1
    -\frac{p_2\cdot p_4}{m_2^2}
    m_1^2B_1
    {\Big \}}.
\end{eqnarray}
and $A_i=a_i \eta$, $B_i=b_i
    \eta$ $(i=1, 2)$.

(3) For $\Lambda_c\rightarrow \Lambda K^+\rightarrow p\pi^0$, we
have similar expressions
\begin{eqnarray}
&&  \non
   A^{FSI}_{\Lam K^{+} \to p \pi^0}
\\    \non
&= &
  \frac{1}{2}\int \frac{d^3 p_1}{(2\pi)^3 2E_1}\frac{d^3 p_2}
    {(2\pi)^3 2E_2}(2\pi)^4 \delta^4(p_1+p_2-p_{\Lamc})
      \cdot A(\Lamc \to \Lam K^{*+})
     \lan \Lam K^{+} |S|p \pi^0 \ran
\\
&= &
    \int^1_{-1}
     \frac{d(cos\theta)|\overrightarrow p_1|}{16\pi M}
     \frac{iF(k^2)g_{K^{*} N\Lam}g_{K^{*+}K\pi}}{k^2-m_{K^{*}}^2}
     {\bar U}_p(p_3)
    i(A H_1^c+B H_2^c \gamma_5)U_{\Lamc} (p_{\Lamc}),
\end{eqnarray}
where $A, B$ take the values in Eq.(\ref{ab2}) and $H_1^c, H_2^c$
are
\begin{eqnarray}
H_1^c &= &  \non
    {\Big [}
    -4p_{\Lamc}\cdot p_1+m_1^2
    -\frac{2\kappa^{\prime}_{\rho}}{m_N+m_\Lam}
    (m_1^2m_3-m_1^2M-2p_{\Lamc}\cdot p_1 m_3)
    -\frac{m_1^2(m_2^2-m_4^2)}{m^2_{K^{*}}}
    {\Big ]}
\\  \non
& &
    +m_1
    {\Big [}
    -2M+m_3
    +\frac{2\kappa^{\prime}_{\rho}}{m_N+m_\Lam}(p_2\cdot p_4-M^2
    +2p_{\Lamc}\cdot p_1+m_3M)
    +\frac{(m_2^2-m_4^2)m_3}{m^2_{K^{*}}}
    {\Big ]},
\\    \non
H_2^c &= &
    {\Big [}
    -4p_{\Lamc}\cdot p_1+m_1^2
    -\frac{2\kappa^{\prime}_{\rho}}{m_N+m_\Lam}
    (m_1^2m_3+m_1^2M-2p_{\Lamc}\cdot p_1 m_3)
    -\frac{m_1^2(m_2^2-m_4^2)}{m^2_{K^{*}}}
    {\Big ]}
\\    \non
& &
    +m_1
    {\Big [}
    2M+m_3
    +\frac{2\kappa^{\prime}_{\rho}}{m_N+m_\Lam}(p_2\cdot p_4-M^2+
    2p_{\Lamc}\cdot p_1-m_3M)
    +\frac{(m_2^2-m_4^2)m_3}{m^2_{K^{*}}}
    {\Big ]},
\end{eqnarray}
where $\kappa^{\prime}_{\rho}$ is from the interaction vertex of
(\ref{6aa}) and we have $p_1=p_{\Lam}, p_2=p_{K^{*+}}, p_3=p_p$,
$p_4=p_{\pi^0}$.\\

(4) For $\Lambda_c\rightarrow \Lambda K^{*+}\rightarrow p\pi^0$,
\begin{eqnarray}
\label{8b} &&  \non
   A^{FSI}_{\Lam K^{*+} \to p \pi^0}
\\    \non
&= &
  \frac{1}{2}\int \frac{d^3 p_1}{(2\pi)^3 2E_1}\frac{d^3 p_2}
    {(2\pi)^3 2E_2}(2\pi)^4 \delta^4(p_1+p_2-p_{\Lamc})
      \cdot A(\Lamc \to \Lam K^{*+})
     \lan \Lam K^{*+} |S|p \pi^0 \ran
\\
&= &
   \int^1_{-1} \frac{d(cos\theta)|\overrightarrow p_1|}{16\pi M}
   \frac{iF(k^2)g_{K\Lam N}g_{K^{*+}K \pi}}{k^2-m_{K}^2}
   i{\bar U}_p(p_3)
   [A_d+B_d\gamma_5]
   U_{\Lamc},
\end{eqnarray}
where $p_1=p_{\Lam}, p_2=p_{K^{*+}}, p_3=p_p$ and $p_4=p_{\pi^0}$.
$A_d, B_d$ are defined as
\begin{eqnarray}
\non A_d &= &
    2m_1
    {\Big \{}
     A_1(M-m_3)
    +A_2p_1\cdot (p_3-p_{\Lamc})
    +\frac{p_2\cdot p_4}{m_2^2}
    {\Big [}
   -A_1 M
   +A_2p_1\cdot (p_{\Lamc}-p_1)
    {\Big ]}
    {\Big \}}
\\   \non
& &
    +2
    {\Big \{}
    2p_1\cdot p_3A_1
    -\frac{p_2\cdot p_4}{m_2^2}
    m_1^2A_1
    {\Big \}},
\\  \non B_d
&= &
    2m_1
    {\Big \{}
     B_1(-M-m_3)
    +B_2p_1\cdot (p_3-p_{\Lamc})
    +\frac{p_2\cdot p_4}{m_2^2}
    {\Big [}
    B_1 M
   +B_2p_1\cdot (p_{\Lamc}-p_1)
    {\Big ]}
    {\Big \}}
\\   \non
& &
    +2
    {\Big \{}
     2p_1\cdot p_3B_1
    -\frac{p_2\cdot p_4}{m_2^2}
    m_1^2B_1
    {\Big \}},
\end{eqnarray}
and $A_i=a_i \tilde{\eta}$, $B_i=b_i
    \tilde{\eta}$ $(i=1, 2)$.

\end{document}